\documentclass{elsart}
\usepackage{amssymb,amsfonts,amsmath,graphicx,subfigure}
\usepackage{graphics,graphicx,epsfig}
\begin{document}
\begin{frontmatter}
\title{Clustering stock market companies via chaotic map synchronization}

\author[a]{N. Basalto}, 
\author[b,c]{R. Bellotti}, 
\author[b]{F. De Carlo\corauthref{e}}\ead{francesco.decarlo@ba.infn.it}
\corauth[e]{Corresponding author. Tel. +39-0805442364},
\author[b]{P. Facchi}, 
\author[b]{S. Pascazio} 

\address[a]{Institute for Advanced Studies at University of Pavia,\\
via Bassi 6, I-27100 Pavia, Italy.}
\address[b]{Dipartimento di Fisica, Universit\`a di Bari, and Istituto Nazionale di Fisica Nucleare, Sezione di
Bari, via Amendola 173, I-70126 Bari, Italy.}
\address[c]{TIRES, Center of Innovative Technologies for Signal Detection and Processing,\\
via Amendola 173, I-70126 Bari, Italy.}

\begin{abstract}
A pairwise clustering approach is applied to the analysis of the
Dow Jones index companies, in order to identify similar temporal
behavior of the traded stock prices. To this end, the chaotic map
clustering algorithm is used, where a map is associated to each
company and the correlation coefficients of the financial time
series are associated to the coupling strengths between maps. The
simulation of a chaotic map dynamics gives rise to a natural
partition of the data, as companies belonging to the same
industrial branch are often grouped together. The identification
of clusters of companies of a given stock market index can be
exploited in the portfolio optimization strategies.
\end{abstract}

\end{frontmatter}

\section{Introduction}
\label{Introduction}

Stock markets are recently triggering a growing interest in the
physicists' community. The objective of this attention is to
understand the underlying dynamics which rules the companies'
stock prices. In particular, it would be useful to find, inside a
given stock market index, groups of companies sharing similar
temporal behavior. To this purpose, a clustering approach to the
problem may represent a good strategy. Clustering deals with the
partitioning of a set of \textit{N} elements into \textit{K}
clusters, based on a suitable (and not unique) similarity
criterion \cite{fukunaga}. Non-parametric methods represent the
optimal strategy when a hierarchical structure, rather than a
fixed partition, of the data should be obtained: this is the case
with stock index dynamics and portfolio optimization strategies
\cite{elton,bouchaud}. Examples of non-parametric methods are the
\textit{linkage} (\textit{agglomerative} and \textit{divisive})
algorithms \cite{jain}, whose output is a \textit{dendrogram}
displaying the full hierarchy of the clustering solutions at
different scales. The agglomerative approaches merge, at each
step, the two clusters with the smallest \textit{distance},
starting from clusters containing only one element. Here we use a
non-parametric clustering approach, named chaotic map clustering
(CMC) \cite{angelini}, which relies on the synchronization
properties of a chaotic map system \cite{kaneko,manrubia} to
obtain a hierarchy of classes without any assumptions on the
underlying structure of the data.
\\
This paper is organized as follows: in section \ref{pcmc} we give
a brief review of the chaotic map algorithm, suitably modified for
pairwise clustering of financial times series. Section \ref{fts}
deals with the analysis of the companies' stock prices. Finally,
some conclusions are drawn in section \ref{conclusions}.

\section{Pairwise chaotic map clustering}
\label{pcmc}

Chaotic map clustering has been introduced as a central algorithm,
where the elements to cluster are embedded in a
\textit{D}-dimensional feature space. In such a picture, the
data-points are viewed as sites of a grid, hosting a chaotic map
dynamics: the map variables $x_{i}\in[-1,1]$, $i=1,\dots,N$, are
assigned to each site of the lattice, and short-range interactions
between neighboring maps are introduced as exponential decreasing
function of the site distance. In the stationary regime, clusters
of synchronized maps appear, corresponding to high density regions
in the original data space. The mutual information between maps is
used both as a similarity index for building the clusters, and a
scale parameter for reconstructing the hierarchical tree.
\\
It should be remarked that a pairwise version of the algorithm can
be easily implemented if an $N\times N$ matrix of similarities
(not necessarily distances in the mathematical sense) is provided
instead of the feature vectors for all data.
\\
As far as one deals with clustering temporal patterns $y_{i}(t)$,
the correlation coefficients $c_{ij}\in[-1,1]$ seem to be a
natural measure of similarity:
\begin{equation} c_{ij}=\frac{\langle Y_{i}Y_{j}\rangle - \langle
Y_{i}\rangle\langle Y_{j}\rangle}{\sqrt{(\langle Y_{i}^{2}\rangle
- \langle Y_{i}\rangle^{2})(\langle Y_{j}^{2}\rangle - \langle
Y_{j}\rangle^{2})}}~, \label{corr}
\end{equation}
where the temporal averages are computed over the time series
length. In \cite{kullmann}, the correlation coefficients between
financial time series are used as entries into the
super-paramagnetic clustering (SPC) algorithm \cite{SPC,domany}.
The SPC algorithm shares the same philosophy of the CMC approach,
the physical system used to partition the data being an
inhomogeneous ferromagnetic model: Potts spin $s_{i}$ are
assigned, instead of map variables, to each data-point and
short-range interactions between neighboring sites are introduced.
The spin-spin correlation function replaces the mutual information
as similarity index for clustering data. In the super-paramagnetic
regime, domains of aligned spins appear, corresponding to the
classes present in the data.
\\
Kullmann et al. generalize the SPC to the case of
anti-ferromagnetic couplings by introducing the following
spin-spin strength as a function of the correlation coefficients
$c_{ij}$ \cite{kullmann}:
\begin{equation}
J_{ij}=\textrm{sgn}(c_{ij})\Big(1-\exp\Big\{-\frac{n-1}{n}\Big[\frac{c_{ij}}{a}
\Big]^{n}\Big\} \Big)~, \label{J}
\end{equation}
where the sign function \textrm{sgn} maps positive/negative
correlations between companies' stock prices into
positive/negative interactions between Potts spins, \textit{n} is
an even positive integer tuning the shape of the interaction
function (whose value should be chosen so that a stable
non-trivial partition can be obtained inside the hierarchical
solution), and \textit{a} is the average of the largest
correlation coefficients for each sequence \cite{kullmann}:
\begin{equation}
a=\frac{1}{N}\sum_{i=1}^{N}\max_{j}(c_{ij})~. \label{aa}
\end{equation}

We shall try to follow a similar strategy in our CMC approach. We
first observe that, in order to implement a chaotic map dynamics,
the correlation coefficients between financial time series should
be mapped into positive interactions between maps, ranging in
[$0,1$]. Hence, we are naturally led to adopt the couplings
(\ref{J}) for $c_{ij}\geq 0$, while setting $J_{ij}=0$ for
$c_{ij}<0$. In this way, we build up a partially coupled map
lattice with exponential increasing interactions between
positively correlated companies. In the case of randomly coupled
systems, although exact synchronization and formation of clusters
of identical dynamical states are not found as in the globally
coupled case \cite{kaneko}, yet, clusters of \textit{almost}
synchronized maps are still observed, even for a significant
fraction (up to $40-45~\%$) of lacking connections
\cite{manrubia}. By retaining the interactions only between
positively correlated time series, we bias the formation of
\textit{almost} synchronized maps to correspond to groups of
companies sharing the same temporal behavior, while
anti-correlated companies are likely to belong to different
clusters. The chaotic map dynamics reads
\begin{equation}
x_i (\tau+1) = \frac{1}{C_i}\sum_{j\ne i} J_{ij}
f\left(x_j(\tau)\right) ~, \label{x}
\end{equation}
where $f(x)=1-2x^2$ is the logistic map, $C_i =\sum_{j\ne i}
J_{ij}$ is a normalization factor, and $\tau$ denotes the
evolution time of the chaotic map system (not to be confused with
the real time \textit{t} of the financial series). A detailed
description of the above mentioned dynamics for clustering
purposes is described elsewhere \cite{angelini}; roughly speaking,
after a certain equilibration time, the dynamics (\ref{x}) yields
a partition of the maps $x_i$ into synchronized clusters, that
remain stable during the remaining part of the $\tau$-evolution.
Applications of the CMC algorithm cover a number of fields, such
as buried land-mines detection by dynamic infrared imaging
\cite{marangi1,marangi2}, human evolution study with mitochondrial
\textit{DNA} sequences \cite{marangi3}, and diagnosis of
pathological electroencephalographic patterns affected by
Huntington's disease \cite{bellotti1,bellotti2}.

\section{Application to financial time series}
\label{fts}

Here we apply the CMC algorithm to cluster the companies of the
Dow Jones (DJ) market index, including $N=30$ stocks, whose names
are listed in Appendix \ref{appendix}, together with the
identifying tickers and the related industrial branches. We first
analyze one-year time periods, from 1998 to 2002. For each year,
the correlation coefficients (\ref{corr}) are computed for the
logarithmic daily price variation time series
\begin{equation}\label{ts}
    y_{i}(t) = \ln(P_{i}(t+1)) - \ln(P_{i}(t)) ~,
\end{equation}
where $P_{i}(t)$ is the closure price of stock \textit{i} at day
\textit{t}.
\\
It should be remarked that, for each investigated period, the
number of pairs of anti-correlated companies $N_{c<0}$ is very
small in comparison with the total number of pairs $N(N-1)/2=435$,
and the mean value of the anticorrelations $\langle
c\rangle_{c<0}$ is almost zero (see table 1). At this point it
should be stressed that the very fact that all stocks are
correlated, and practically lack any anticorrelation, make not
easy any possible clustering procedure.
\\
As a result of the processing, a dendrogram displays the
hierarchical structure of the clusters at different values of the
mutual information $I_{ij}$ defined as follows:
\begin{itemize}
    \item extract a bitwise sequence $S_i$ from each map
    $x_{i}(t)$, such that
    \begin{equation} S_{i}=\left\{%
    \begin{array}{ll}
    1, & \hbox{~if $x_i(t)\geq 0$;} \\
    0, & \hbox{~otherwise;} \\
    \end{array}%
    \right. \end{equation}
    \item evaluate the probability $P(S_i)$ as the number of times the state
    $S_{i}$ occurs along the sequence $S_{i}$, normalized to the sequence length;
    in a similar way, $P(S_i,S_j)$ is the frequency of simultaneous occurrence of the
    states
    $(S_{i},S_{j})$ along the sequences $S_{i}$ and $S_{j}$;
    \item compute the string entropy $H_{i}$ and the joint entropy
    $H_{ij}$ as
\begin{equation}
H_i = -\sum_{S_i=0,1}P(S_i)\ln P(S_i) ~, \label{Boltzmann-entropy}
\end{equation}
\begin{equation}
H_{ij} = -\sum_{S_i=0,1}\sum_{S_j=0,1}P(S_i,S_j)\ln P(S_i,S_j) ~;
\label{joint-entropy}
\end{equation}
    \item the mutual information is then given by: $I_{ij} = H_{i} + H_{j} -
    H_{ij}$;
\end{itemize}
The mutual information is a measure of the correlations between
maps \cite{sole}, ranging between $I_{ij}=0$, for maps evolving
independently, and $I_{ij}=\ln(2)$, for exactly synchronized maps.
For this reason, $I_{ij}$ can be appropriately adopted as a
similarity index for clustering the companies: by cutting the
dendrogram at a certain level $I\in[0,\ln(2)]$, the clusters thus
obtained are made up of companies whose associated maps are
characterized by $I_{ij}\geq I$. The level \textit{I} can be
suitably chosen by relying on a certain stability criterion of the
clustering solution. To this purpose, the cluster entropy $S(I)$
\cite{kaneko} can be used to select the most stable partition
among the whole hierarchy yielded by the algorithm, by looking for
a plateau in the widest possible range of \textit{I} values:
\begin{equation}\label{clust-ent}
    S(I) = -\sum_{k=1}^{N_{I}}P_{I}(k)\ln P_{I}(k) ~,
\end{equation}
where $P_{I}(k)$ is the fraction of elements belonging to cluster
\textit{k}, and $N_{I}$ is the number of clusters found at level
\textit{I}.
\\
This model depends on one parameter, the positive even integer
number \textit{n}, which tunes the range of the interactions
(\ref{J}). For each period, the optimal value of the parameter
\textit{n} should be chosen according to the stability criterion
of the entropy (\ref{clust-ent}), at different cluster partitions.
As an example, we consider the processing relative to the year
1999: figure 1 displays the entropy \textit{S} in the plane
spanned by \textit{I} (mutual information) and \textit{n}, with
$n=2,4,6,\ldots,24$. We choose $n=8$ to be the optimal value, by
looking for the widest range of constant values of \textit{S},
along the \textit{I}-direction ($0.4\lesssim I \lesssim 0.6$).
\\
Once this parameter has been adjusted, the full hierarchy of
clusters can be displayed by a dendrogram: figure 2 shows the
result obtained for the year 1999. The dendrogram has been cut in
the region of stable partitions at $I\simeq 0.6$. For low value of
mutual information, all pairs of companies are linked together in
one single cluster, which splits into two big clusters at
$I=0.16$: on one side, we clearly recognize companies dealing
mainly with capital goods (BA, CAT, HON) and basic materials (AA,
DD, IP). On the other side, we find a cluster of strongly
correlated companies represented by the branch marked by a star.
This cluster, which gradually breaks as the mutual information
approaches its maximum value $I=\ln(2)$, groups together different
industrial branches: financial (C, AXP, JPM), services (DIS, HD,
MCD, SBC, T, WMT), healthcare (JNJ, MRK), conglomerates (GE, UTX),
consumer non-cyclical (GM, KO, MO, PG). Besides this cluster, it
should be remarked the formation of technological cores (IBM and
HPQ, INTC and MSFT).
\\
This analysis has been carried out for each of the 5 years
considered (1998-2002). In the following, we report the main
clusters found for different years, together with the values
chosen for the parameter \textit{n}, and the values of the mutual
information at which the dendrogram has been cut. Sub-clusters of
companies belonging to the same industrial branch have been
underbraced:
\begin{itemize}
    \item \textbf{Year 1998, $n=16$, $I=0.62$}
    \begin{description}
        \item[(1)] $\underbrace{\textrm{DIS ~ MCD ~ T ~ WMT}}$ ~ $\underbrace{\textrm{KO ~ MO ~ PG}}$ ~ $\underbrace{\textrm{JNJ ~ MRK}}$
        \item[(2)] $\underbrace{\textrm{AXP ~ C ~ JPM}}$ ~ GM
    \end{description}
    \item \textbf{Year 1999, $n=8$, $I=0.24$}
    \begin{description}
        \item[(1)] $\underbrace{\textrm{DIS ~ HD ~ MCD ~ SBC ~ T ~ WMT}}$ ~ $\underbrace{\textrm{KO ~ MO ~ PG}}$ ~
        $\underbrace{\textrm{AXP ~ C ~ JPM}}$ ~ $\underbrace{\textrm{JNJ ~ MRK}}$ ~
        $\underbrace{\textrm{INTC ~ MSFT}}$ ~ $\underbrace{\textrm{GE ~ UTX}}$ ~ GM
        \item[(2)] $\underbrace{\textrm{BA ~ CAT ~ HON}}$ ~ $\underbrace{\textrm{DD ~ IP}}$ ~ EK ~ XOM
    \end{description}
    \item \textbf{Year 2000, $n=18$, $I=0.26$}
    \begin{description}
        \item[(1)] $\underbrace{\textrm{BA ~ CAT ~ HON}}$ ~ $\underbrace{\textrm{AA ~ DD ~ IP}}$ ~ $\underbrace{\textrm{KO ~ PG}}$ ~
        $\underbrace{\textrm{MMM ~ UTX}}$ ~ EK ~ MCD
        \item[(2)] $\underbrace{\textrm{AXP ~ C ~ JPM}}$ ~ $\underbrace{\textrm{SBC ~ T}}$ ~ GE ~
        GM
    \end{description}
    \item \textbf{Year 2001, $n=20$, $I=0.15$}
    \begin{description}
        \item[(1)] $\underbrace{\textrm{DIS ~ HD ~ MCD ~ SBC ~ T ~ WMT}}$ ~ $\underbrace{\textrm{BA ~ CAT ~ HON}}$ ~ $\underbrace{\textrm{AXP ~ C ~ JPM}}$ ~ $\underbrace{\textrm{AA ~ DD ~ IP}}$ ~
        $\underbrace{\textrm{GE ~ MMM ~ UTX}}$ ~ $\underbrace{\textrm{EK ~ GM}}$ ~ MO ~ XOM
        \item[(2)] $\underbrace{\textrm{HPQ ~ IBM ~ INTC ~ MSFT}}$
    \end{description}
    \item \textbf{Year 2002, $n=16$, $I=0.62$}
    \begin{description}
        \item[(1)] $\underbrace{\textrm{AA ~ DD ~ IP}}$ ~ $\underbrace{\textrm{CAT ~ HON}}$ ~ $\underbrace{\textrm{MMM ~ UTX}}$ ~ GM ~ MCD ~ XOM;
        \item[(2)] $\underbrace{\textrm{AXP ~ C ~ JPM}}$ ~ $\underbrace{\textrm{DIS ~ SBC}}$ ~ EK ~ GE ~ MRK
        \item[(3)] $\underbrace{\textrm{HPQ ~ IBM ~ MSFT}}$ ~ HD~.
    \end{description}
\end{itemize}
It is worth stressing the presence of some cores of companies
which remain strongly linked together over periods longer than one
year: financial companies (AXP, C, JPM, 98-02), services (DIS,
MCD, T, WMT, 98-99, 01), consumer non-cyclical (KO, MO, PG,
98-99), basic materials (AA, DD, IP, 00-02), capital goods (BA,
CAT, HON, 99-01), technology (HPQ, IBM, MSFT, 01-02), healthcare
(JNJ, MRK, 98-99), conglomerates (MMM, UTX, 00-02).
\\
Once a partition of companies has been obtained, an efficient
portfolio could be made of one ``representative" stock per
cluster, thus ensuring a diversification of the investment. The
choice of the period length for computing the correlation
coefficients should be related to the flexibility of the
portfolio. From this point of view, an analysis covering the whole
5-year period should be based on more stable correlation
coefficients, thus leading to more stable partitions (i.e., less
hazardous investment), at the cost of a less flexible portfolio.
In figure \ref{dendro-all}, we report the full hierarchy of
clusters found from the whole 5-year length time period ($n=18$).
We want to remark that no anticorrelations have been found for
such period. The main branches of the dendrogram have been marked
by the industrial areas of the companies they are made of.

\section{Conclusions}
\label{conclusions}

In the present work, a pairwise version of the chaotic map
algorithm has been applied to the analysis of the companies'
stocks belonging to the Dow Jones market index. The correlation
coefficients between financial time series have been used as
similarity measures to cluster the temporal patterns. Once the
coupling interactions between maps are taken to be functions of
this feature, the dynamics of such a system leads to the formation
of clusters of companies that can often be identified as different
industrial branches. The clustering output can be exploited to
optimize the portfolio composition.

\begin{table}[p]
\begin{tabular}{|c|c|c|c|c|c|}
  \hline
  Year & 1998 & 1999 & 2000 & 2001 & 2002\\
  \hline
  $N_{c<0}$ & 0 & 25 & 34 & 11 & 1\\
  \hline
  $\langle c \rangle_{c<0}$ & 0 & -0.0453 & -0.0494 & -0.0495 & -0.0071\\
  \hline
\end{tabular}
\label{tab1} \caption{Number of pairs of anti-correlated stocks
$N_{c<0}$ and mean value of the anticorrelation $\langle
c\rangle_{c<0}$. $N_{c<0}$ and $\langle c\rangle_{c<0}$ must be
compared with the total number of pairs $N(N-1)/2=435$ and with
the mean correlation $\langle c\rangle \simeq 0.28$,
respectively.}
\end{table}

\begin{figure}[p]
\begin{center}
\includegraphics*[scale=0.80]{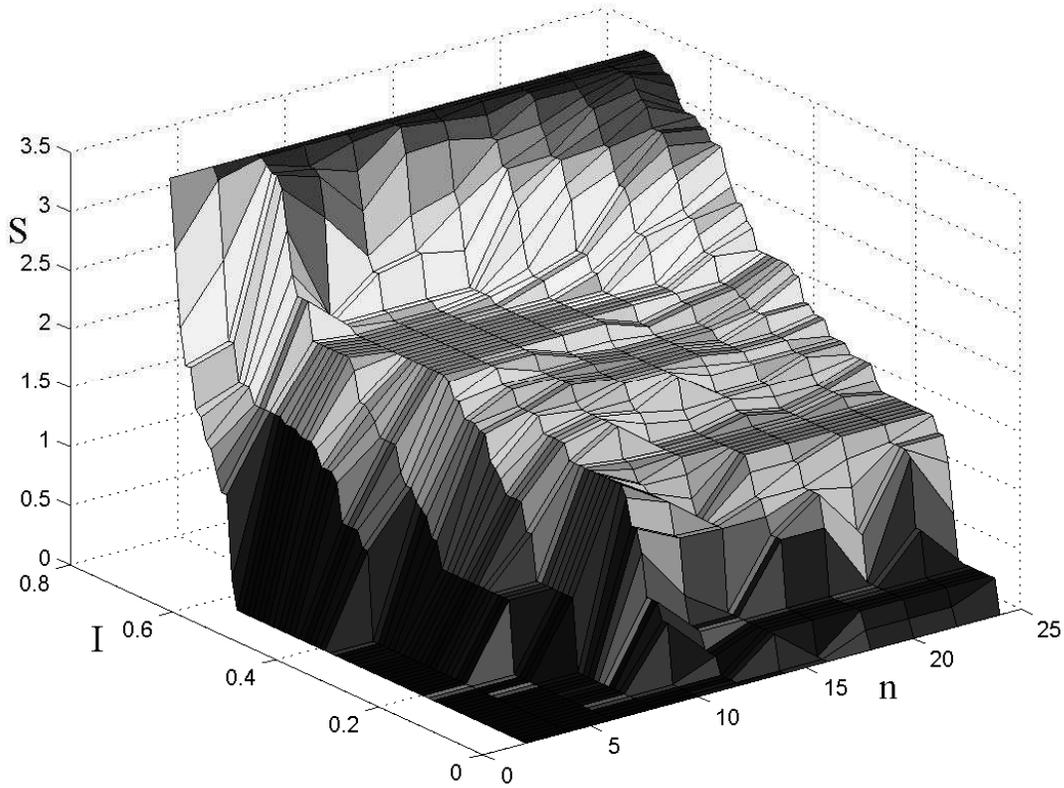}
\caption{Cluster entropy \textit{S} in the plane spanned by the
mutual information \textit{I} and the parameter \textit{n}. The
widest $S$-plateau along the \textit{I}-direction (namely, the
range of values of $I$ for which $S$ is constant) is $0.4\lesssim
I \lesssim 0.6$ and corresponds to $n=8$. This analysis refers to
year 1999.}
\end{center}
\label{entropy}
\end{figure}

\begin{figure}[p]
\begin{center}
\includegraphics*[scale=0.80]{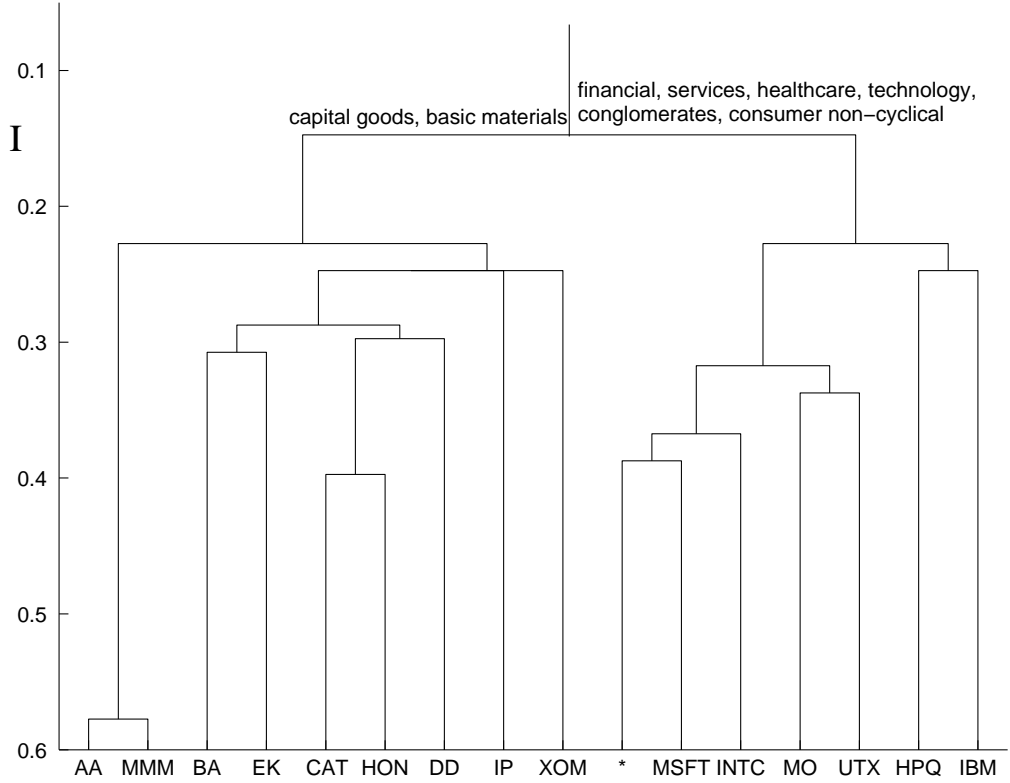}
\caption{Dendrogram obtained for the year 1999 ($n=8$), cut in the
region of stable partitions at $I\simeq 0.6$. The branch marked by
a star (not explicitly shown) groups together different industrial
sub-classes: financial (C, AXP, JPM), services (DIS, HD, MCD, SBC,
T, WMT), healthcare (JNJ, MRK), conglomerates (GE, UTX), consumer
non-cyclical (GM, KO, MO, PG).} \label{dendro_99}
\end{center}
\end{figure}

\begin{figure}[p]
\begin{center}
\includegraphics*[scale=0.80]{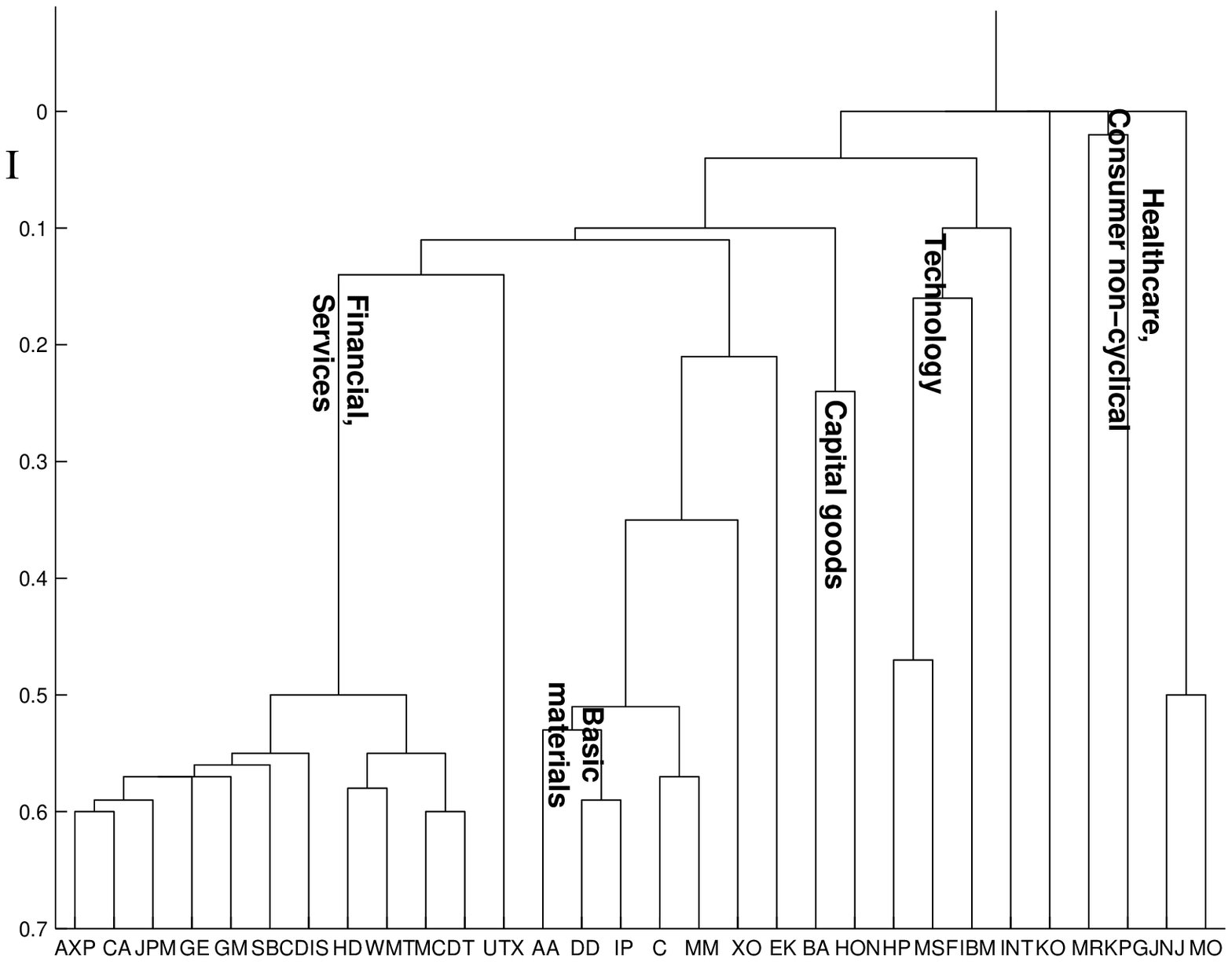}
\caption{Dendrogram found from the whole 5-year time period
1998-2002, with $n=18$. The main branches have been marked by the
industrial areas of the companies they are made of.}
\label{dendro-all}
\end{center}
\end{figure}

\newpage

\appendix
\section{Dow Jones stock market companies} \label{appendix}
\begin{description}
    \item[AA:] Alcoa Inc. - Basic Materials
    \item[AXP:] American Express Co. - Financial
    \item[BA:] Boeing - Capital Goods
    \item[C:] Citigroup - Financial
    \item[CAT:] Caterpillar - Capital Goods
    \item[DD:] DuPont - Basic Materials
    \item[DIS:] Walt Disney - Services
    \item[EK:] Eastman Kodak - Consumer Cyclical
    \item[GE:] General Electrics - Conglomerates
    \item[GM:] General Motors - Consumer Cyclical
    \item[HD:] Home Depot - Services
    \item[HON:] Honeywell International - Capital Goods
    \item[HPQ:] Hewlett-Packard - Technology
    \item[IBM:] International Business Machine - Technology
    \item[INTC:] Intel Corporation - Technology
    \item[IP:] International Paper - Basic Materials
    \item[JNJ:] Johnson \& Johnson - Healthcare
    \item[JPM:] JP Morgan Chase - Financial
    \item[KO:] Coca Cola Inc. - Consumer Non-Cyclical
    \item[MCD:] McDonalds Corp. - Services
    \item[MMM:] Minnesota Mining - Conglomerates
    \item[MO:] Philip Morris - Consumer Non-Cyclical
    \item[MRK:] Merck \& Co. - Healthcare
    \item[MSFT:] Microsoft - Technology
    \item[PG:] Procter \& Gamble - Consumer Non-Cyclical
    \item[SBC:] SBC Communications - Services
    \item[T:] AT\&T Gamble - Services
    \item[UTX:] United Technology - Conglomerates
    \item[WMT:] Wal-Mart Stores - Services
    \item[XOM:] Exxon Mobil - Energy
\end{description}

\end{document}